\PassOptionsToPackage{unicode}{hyperref}
\PassOptionsToPackage{hyphens}{url}
\documentclass[
]{aiaa-tc}
\usepackage{svg}
\usepackage{svg-extract}
\svgsetup{extract=true,extractpath=/tmp/svg-extract}
\usepackage[noblocks]{authblk}
\usepackage{iftex}
\ifPDFTeX
  \usepackage[T1]{fontenc}
  \usepackage[utf8]{inputenc}
  \usepackage{lmodern}
  \usepackage{textcomp} %
\else %
  \usepackage{unicode-math}
  \defaultfontfeatures{Scale=MatchLowercase}
  \defaultfontfeatures[\rmfamily]{Ligatures=TeX,Scale=1}
\fi
\IfFileExists{upquote.sty}{\usepackage{upquote}}{}
\IfFileExists{microtype.sty}{%
  \usepackage[]{microtype}
  \UseMicrotypeSet[protrusion]{basicmath} %
}{}
\makeatletter
\@ifundefined{KOMAClassName}{%
  \IfFileExists{parskip.sty}{%
    \usepackage{parskip}
  }{%
    \setlength{\parindent}{0pt}
    \setlength{\parskip}{6pt plus 2pt minus 1pt}}
}{%
  \KOMAoptions{parskip=half}}
\makeatother
\usepackage{xcolor}
\IfFileExists{xurl.sty}{\usepackage{xurl}}{} %
\IfFileExists{bookmark.sty}{\usepackage{bookmark}}{\usepackage{hyperref}}
\hypersetup{
  pdftitle={New Methods for Computer Tomography Based Ion Thruster Diagnostics and Simulation},
  hidelinks,
  pdfcreator={LaTeX via pandoc}}
\usepackage{longtable,booktabs,array}
\usepackage{calc} %
\usepackage{etoolbox}
\makeatletter
\patchcmd\longtable{\par}{\if@noskipsec\mbox{}\fi\par}{}{}
\makeatother
\IfFileExists{footnotehyper.sty}{\usepackage{footnotehyper}}{\usepackage{footnote}}
\makesavenoteenv{longtable}
\usepackage{graphicx}
\makeatletter
\def\maxwidth{\ifdim\Gin@nat@width>\linewidth\linewidth\else\Gin@nat@width\fi}
\def\maxheight{\ifdim\Gin@nat@height>\textheight\textheight\else\Gin@nat@height\fi}
\makeatother
\setkeys{Gin}{width=\maxwidth,height=\maxheight,keepaspectratio}
\makeatletter
\def\fps@figure{htbp}
\makeatother
\providecommand{\tightlist}{%
  \setlength{\itemsep}{0pt}\setlength{\parskip}{0pt}}
\NewDocumentCommand\citeproctext{}{}
\NewDocumentCommand\citeproc{mm}{%
  \begingroup\def\citeproctext{#2}\cite{#1}\endgroup}
\makeatletter
 \let\@cite@ofmt\@firstofone
 \def\@biblabel#1{}
 \def\@cite#1#2{{#1\if@tempswa , #2\fi}}
\makeatother
\newlength{\cslhangindent}
\newlength{\csllabelwidth}
\newenvironment{CSLReferences}[2] %
 {\begin{list}{}{%
  \setlength{\itemindent}{0pt}
  \setlength{\leftmargin}{0pt}
  \setlength{\parsep}{0pt}
  \ifodd #1
   \setlength{\leftmargin}{\cslhangindent}
   \setlength{\itemindent}{-1\cslhangindent}
  \fi
  \setlength{\itemsep}{#2\baselineskip}}}
 {\end{list}}
\usepackage{calc}

\newcommand{\CSLLeftMargin}[1]{\parbox[t]{\csllabelwidth}{\strut#1\strut}}
\newcommand{\CSLRightInline}[1]{\parbox[t]{\linewidth - \csllabelwidth}{\strut#1\strut}}

\usepackage{fancyhdr}
\usepackage{iepc2024}
\IEPCsubmissionnumber{469}
\makeatletter
\@ifpackageloaded{subfig}{}{\usepackage{subfig}}
\@ifpackageloaded{caption}{}{\usepackage{caption}}
\AtBeginDocument{%
\renewcommand*\figurename{Figure}
\renewcommand*\tablename{Table}
}
\AtBeginDocument{%

}
\newcounter{pandoccrossref@subfigures@footnote@counter}
\newenvironment{pandoccrossrefsubfigures}{%
\setcounter{pandoccrossref@subfigures@footnote@counter}{0}
\begin{figure}\centering%
\gdef\global@pandoccrossref@subfigures@footnotes{}%
\DeclareRobustCommand{\footnote}[1]{\footnotemark%
\stepcounter{pandoccrossref@subfigures@footnote@counter}%
\ifx\global@pandoccrossref@subfigures@footnotes\empty%
\gdef\global@pandoccrossref@subfigures@footnotes{{##1}}%
\else%
\g@addto@macro\global@pandoccrossref@subfigures@footnotes{, {##1}}%
\fi}}%
{\end{figure}%
\addtocounter{footnote}{-\value{pandoccrossref@subfigures@footnote@counter}}
\@for\f:=\global@pandoccrossref@subfigures@footnotes\do{\stepcounter{footnote}\footnotetext{\f}}%
\gdef\global@pandoccrossref@subfigures@footnotes{}}
\@ifpackageloaded{float}{}{\usepackage{float}}
\floatstyle{ruled}
\@ifundefined{c@chapter}{\newfloat{codelisting}{h}{lop}}{\newfloat{codelisting}{h}{lop}[chapter]}
\floatname{codelisting}{Listing}

\makeatother
\usepackage{acro}
\DeclareAcronym{CGLS}{
short = CGLS,
long = Conjugate Gradient Least Squares
}
\DeclareAcronym{CT}{
short = CT,
long = Computed Tomography
}
\DeclareAcronym{EM}{
short = EM,
long = Expectation-Maximization
}
\DeclareAcronym{EP}{
short = EP,
long = Electric Propulsion
}
\DeclareAcronym{ESA}{
short = ESA,
long = European Space Agency
}
\DeclareAcronym{FBP}{
short = FBP,
long = Filtered Back Projection
}
\DeclareAcronym{FDK}{
short = FDK,
long = Feldkamp-Davis-Kress
}
\DeclareAcronym{GSTP}{
short = GSTP,
long = General Support Technology Programme
}
\DeclareAcronym{IEPC}{
short = IEPC,
long = International Electric Propulsion Conference
}
\DeclareAcronym{MAR}{
short = MAR,
long = Metal Artifact Reduction
}
\DeclareAcronym{MBIR}{
short = MBIR,
long = Model-Based Iterative Reconstruction
}
\DeclareAcronym{SART}{
short = SART,
long = Simultaneous Algebraic Reconstruction Technique
}
\DeclareAcronym{SIRT}{
short = SIRT,
long = Simultaneous Iterative Reconstruction Technique
}
\DeclareAcronym{SVMBIR}{
short = SVMBIR,
long = Super-Voxel Model-Based Iterative Reconstruction
}
\DeclareAcronym{µCT}{
short = µCT,
long = Microscopic Computed Tomography Device
}
\ifLuaTeX
  \usepackage{selnolig}  %
\fi

\title{New Methods for Computer Tomography Based Ion Thruster
Diagnostics and Simulation}

\author{Jörn Krenzer\footnote{%
Research Assistant, %
Institute for Physics, %
Laboratory for Plasma Technology, %
joern.krenzer@unibw.de%
}%
}
\author{Felix Reichenbach\footnote{%
Student Assistant, %
Institute for Physics, %
Laboratory for Plasma Technology, %
felix.reichenbach@unibw.de%
}%
}
\author{Jochen Schein\footnote{%
Chairholder, %
Institute for Physics, %
Laboratory for Plasma Technology, %
jochen.schein@unibw.de%
}%
}
\affil{\normalsize\it{Bundeswehr University
Munich, Werner-Heisenberg-Weg 39, 85577 Neubiberg, Germany }}

\date{2022-06-04}

\begin{document}
\maketitle

\printacronyms[name=Acronyms and Abbreviations]

\section{Introduction}\label{introduction}

\subsection{Motivation}\label{motivation}

Non-destructive X-ray imaging of thruster parts and assemblies down to
the scale of several micrometers is a key technology for electric
propulsion research and engineering. It allows for thorough product
assurance, rapid state acquisition and implementation of more detailed
simulation models to understand the physics of device wear and erosion.

Being able to inspect parts as 3D density maps allows insight into inner
structures hidden from observation. Generating these density maps and
also constructing three dimensional mesh objects for further processing
depends on the achievable quality of the reconstruction, which is the
inverse of Radon's transformation connecting a stack of projections
taken from different angles to the original object's structure.

Reconstruction is currently flawed by strong mathematical artifacts
induced by the many aligned parts and stark density contrasts commonly
found in electric propulsion thrusters.

For the raw image acquisition today's industrial X-ray \acp{µCT} offer
enough performance, but for the step of optimization, mathematical
reconstruction and volume generation no transparent and reviewable
software package is currently accessible to the electric propulsion
community which is able to tackle the specific properties and
requirements of electric thrusters.

This work shows the a comparison of classical and modern reconstruction
algorithms which have been investigated to determine the further paths
viable to develop transparent reconstruction solutions for usage in
electric propulsion diagnostics.

\section{Fundamentals}\label{fundamentals}

In this section a brief explanation of the functional principle of CT
and the reconstruction process will be given. To be concise,
mathematical explanations are left out; readers interested in a more
thorough theoretical explanation are recommended to refer to literature
{[}\citeproc{ref-hermanFundamentalsComputerizedTomography2009}{1}{]} or,
for a brief introduction, the previous work
{[}\citeproc{ref-iepc-2022-260}{2}{]} presented at IEPC 2022.

\subsection{X-Ray Tomographic Imaging}\label{x-ray-tomographic-imaging}

\begin{figure}
\centering\includegraphics{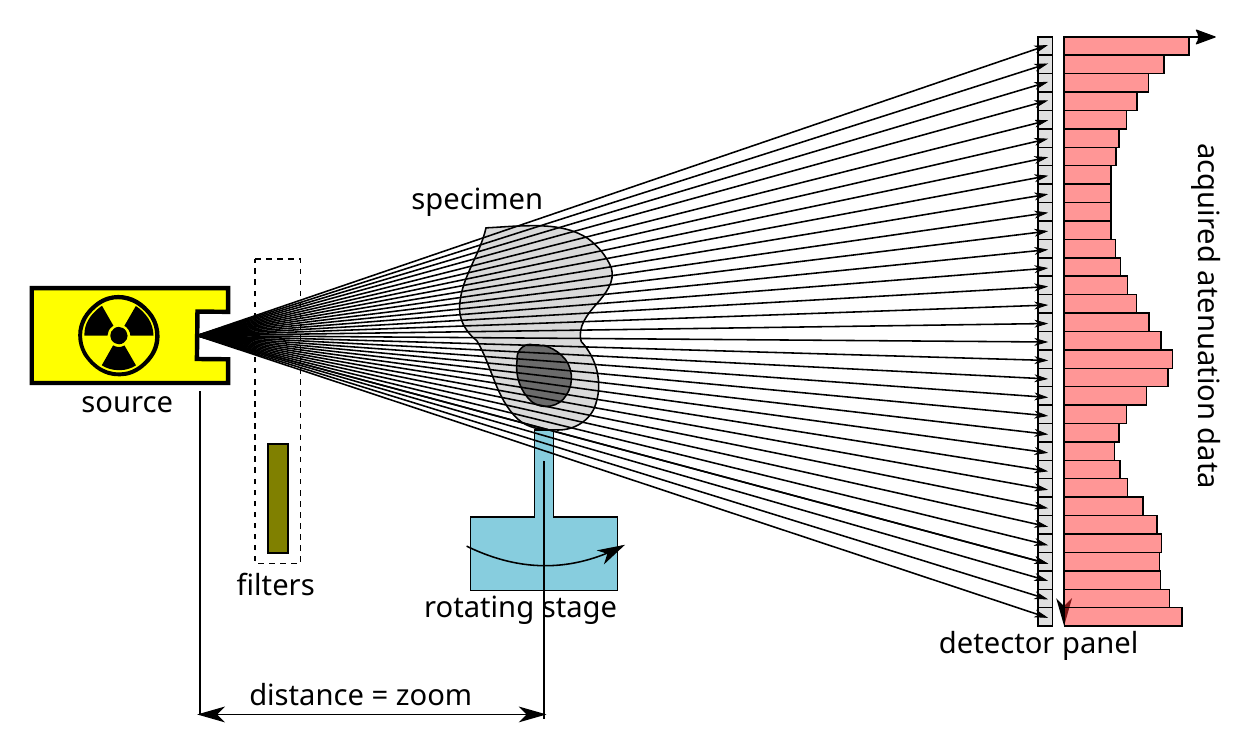}
\caption{Functioning principle schematic of a simple
\ac{µCT}}\label{fig:function-mct}
\end{figure}

As a quick reminder we cover the schematic functional principle of a
\ac{µCT} here. A X-Ray \ac{CT} device acquires attenuation (dampening)
values along rays going from a source through a specimen into a detector
array fig.~\ref{fig:function-mct}. Depending on the setup these
radiographs can be one or two dimensional. By rotating the specimen or
the device, another radiograph is acquired. The stack of these
radiographs is the input for the reconstruction process.

\begin{figure}
\centering\includegraphics{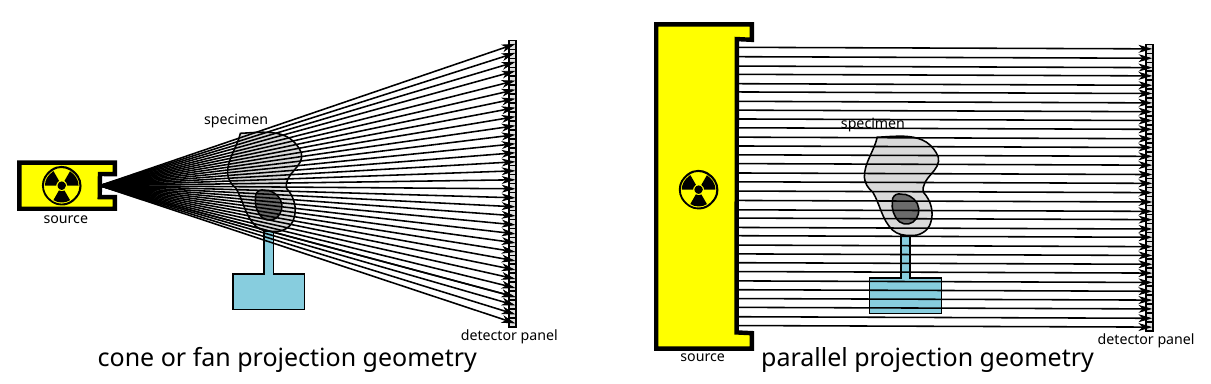}
\caption{Different projection geometries found typically in \ac{CT}
devices. Both classes of geometries can be realized in a two or three
dimensional variants}\label{fig:projection-geometries}
\end{figure}

\ac{CT} devices come with different projection geometries, both two and
three-dimensional fig.~\ref{fig:projection-geometries}. For \acp{µCT},
which are best suited for electric propulsion diagnostics the geometry
is customary a flat-fan or flat-cone geometry. Flat refers to the
detector not being curved towards the source but flat. The projection
geometry is crucial information for the reconstruction process and not
all algorithms are able to process data acquired in all the possible
projection geometries.

\subsection{Reconstruction Process}\label{reconstruction-process}

\begin{figure}
\centering\includegraphics[width=\textwidth,height=15cm]{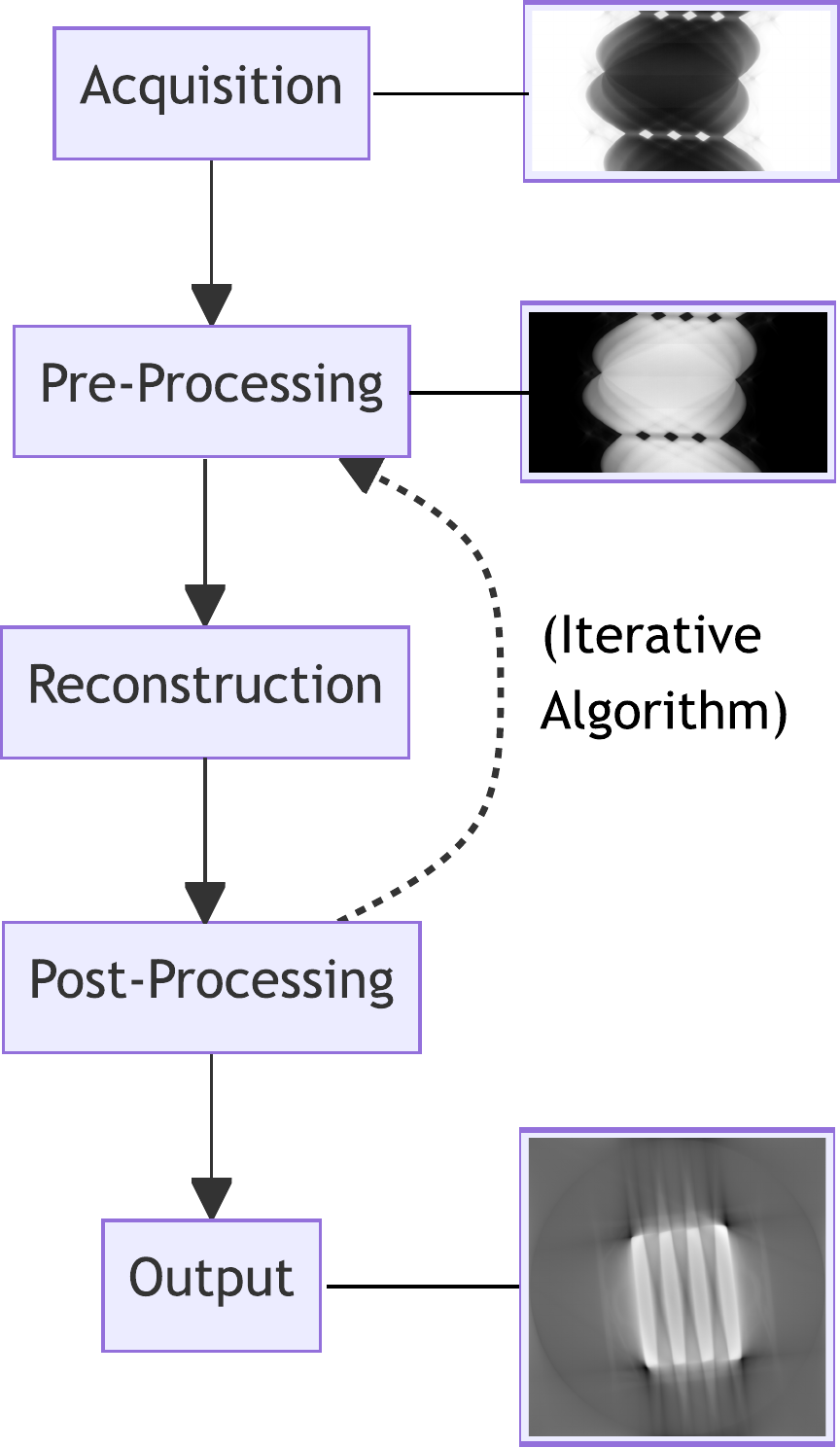}
\caption{Flow of the reconstruction process. Radiograph data is acquired
or loaded, preprocessing yields sinograph. Reconstruction can be
iterative in some algorithms allowing for enhanced processing.
Postprocessing generates Image stacks for 3 dimensional density
data.}\label{fig:reconstruction-flow}
\end{figure}

The basis of reconstruction is the implementation of an inverse
\textsc{Radon}-tranformation to reverse the integration of attenuation
values along each beam-line as seen in fig.~\ref{fig:function-mct}. One
of the oldest algorithms for this is \ac{FBP} to reverse the projection
process which works on one projection at a time. Newer algorithms
implement iterative approaches, so initial guesses, weights or other
techniques to influence reconstruction can be applied in between steps
and also simultaneous approaches which take into account data from
multiple or all projection layers at once.

As seen in fig.~\ref{fig:reconstruction-flow} the overall process has
several steps of data processing where additional methods can tie in to
influence result quality. For iterative algorithms it is possible to
directly access each reconstruction step and thus they are well suited
for modification.

\section{Experimental Approach}\label{experimental-approach}

The goal of this investigation is the identification of promising
reconstruction methods as part of a larger endeavor to develop a
specialized workflow for imaging electric propulsion systems in
\acp{µCT}. For this purpose different reconstruction algorithms are
provided with the same input data and results are compared subjectively.
As base for the comparison, an earlier reconstruction image using the
\ac{CT} device's included algorithm is to visualize overall improvement
against proprietary software which tailored to the device.

\subsection{Selection of Algorithms}\label{selection-of-algorithms}

The algorithms are chosen for their relevance in practical use and for
their perceived better coping capabilities with metal artifacts. Some
classical algorithms are chosen as they might be susceptible to
artifacts but are in wide-spread use in software included with devices
typically well suited for electric propulsion diagnosis. Investigated
algorithms are:

\begin{itemize}
\tightlist
\item
  \ac{FBP}
  {[}\citeproc{ref-hermanFundamentalsComputerizedTomography2009}{1}{]}

  \begin{itemize}
  \tightlist
  \item
    A two dimensional algorithm with high popularity in micro CT device
    software
  \item
    Low computational cost
  \item
    Said to be prone to strong artifact generation
  \item
    Quality dependent on the filter which is used
  \end{itemize}
\item
  \ac{FDK}
  {[}\citeproc{ref-feldkampPracticalConebeamAlgorithm1984}{3}{]}

  \begin{itemize}
  \tightlist
  \item
    A three dimensional algorithm with high popularity in micro CT
    device software
  \item
    Low computational cost
  \item
    Said to be prone to strong artifact generation
  \end{itemize}
\item
  \ac{SIRT}
  {[}\citeproc{ref-andersenSimultaneousAlgebraicReconstruction1984}{4}{]}

  \begin{itemize}
  \tightlist
  \item
    Available as two and three dimensional variant
  \item
    Iterative algorithm, suitable for iterative artifact reduction
    techniques
  \item
    Medium to high computational cost
  \item
    Generates smooth pictures at the cost of global optimization
  \item
    Better perceived performance with artifacts then \ac{FBP} and
    \ac{FDK}
  \end{itemize}
\item
  \ac{SART}
  {[}\citeproc{ref-andersenSimultaneousAlgebraicReconstruction1984}{4}{]}

  \begin{itemize}
  \tightlist
  \item
    Available as two and three dimensional variant
  \item
    Iterative algorithm, suitable for iterative artifact reduction
    techniques
  \item
    Medium computational cost
  \item
    Generates somewhat noisy pictures but has local optimization
  \item
    Better perceived performance with artifacts then \ac{FBP} and
    \ac{FDK}
  \end{itemize}
\item
  \ac{CGLS}
  {[}\citeproc{ref-hermanFundamentalsComputerizedTomography2009}{1}{]}

  \begin{itemize}
  \tightlist
  \item
    Available as two and three dimensional variant
  \item
    Classical, wide-spread algorithm
  \item
    Low to medium computational cost
  \item
    Better perceived performance with artifacts then \ac{FBP} and
    \ac{FDK}
  \end{itemize}
\item
  \ac{EM} {[}\citeproc{ref-langeEMReconstructionAlgorithms1984}{5}{]}

  \begin{itemize}
  \tightlist
  \item
    A two dimensional, statistical algorithm
  \item
    High computational cost
  \item
    Said to be robust against artifacts at the cost of being highly
    depended on a good a priori guess of specimen geometry
  \end{itemize}
\end{itemize}

\subsection{Test Specification}\label{test-specification}

Each algorithm receives the same input data of a phantom which was build
to emphasize the structural challenges for \ac{CT} imaging in a gridded
ion thrusters ion optic. See {[}\citeproc{ref-iepc-2022-260}{2}{]} for
information on the phantom. Input data was, as is customary in many
micro \ac{CT} systems, acquired in a cone-beam projection geometry.

\begin{figure}
\centering\includegraphics{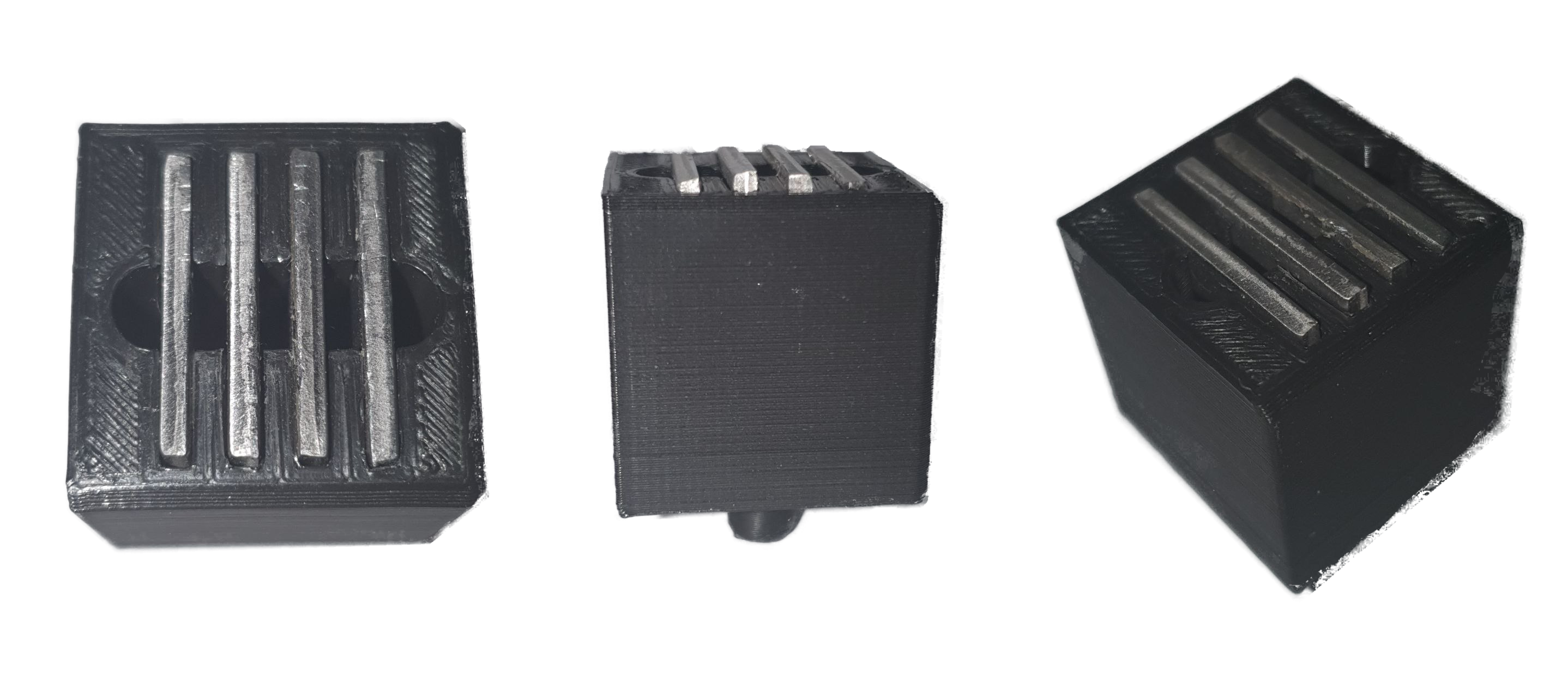}
\caption{Photographs of the phantom used. The structure with four
stainless steel plates in parallel isolated and embedded in plastic
models the simplified structure of an electric propulsion ion
optic}\label{fig:phantom}
\end{figure}

The phantom, by the close proximity of the parallel metal plates,
strongly evokes the metal streaking artifacts and information loss
observed in \ac{CT} survey of gridded ion optics. Thus, any improvement
achieved on imaging of the phantom is likely to hold up to the same
degree in observation of real world ion optic grids.

Testing of the algorithms is conducted by a dedicated software package
written in python 3 to facilitate comparable input and output processes
and thereby minimize the influence of data loading and saving operations
on the perceived results.

To ensure a comparable implementation-quality of the algorithms, the
ASTRA Toolbox was chosen as reconstruction backend
{[}\citeproc{ref-vanaarleASTRAToolboxPlatform2015}{6},\citeproc{ref-aarleFastFlexibleXray2016}{7}{]}.
A comparison with additional algorithms and alternative implementations
was planned using the TomoPy
{[}\citeproc{ref-gursoyTomoPyFrameworkAnalysis2014}{8}{]} and SVMBIR
{[}\citeproc{ref-svmbir-2024}{9}{]} frameworks, but could not be
performed in scope of this work due to the additional integration effort
imposed colliding with time constraints.

\section{Discussion}\label{discussion}

\subsection{Reconstruction Results}\label{reconstruction-results}

In fig.~\ref{fig:results1} the output of all tested algorithms and a
reference picture reconstructed with the \ac{µCT} device-software
(undisclosed algorithm) are presented for comparison.

It should be duly noted, that the pictures are reduced in size and
especially dynamic range as 32 bit floating point values cannot be
rendered in printing or on normal computer screens without down-sampling
the data. These circumstances lead to data-loss cropped to the original
data.

Aside from these necessary transformation not further optimizations,
e.g. beam-hardening-correction, rotation-artifact-reduction etc. have
been applied to assert no interference with achieved quality of
reconstruction.

\begin{figure}
\centering\includegraphics[width=6cm,height=\textheight]{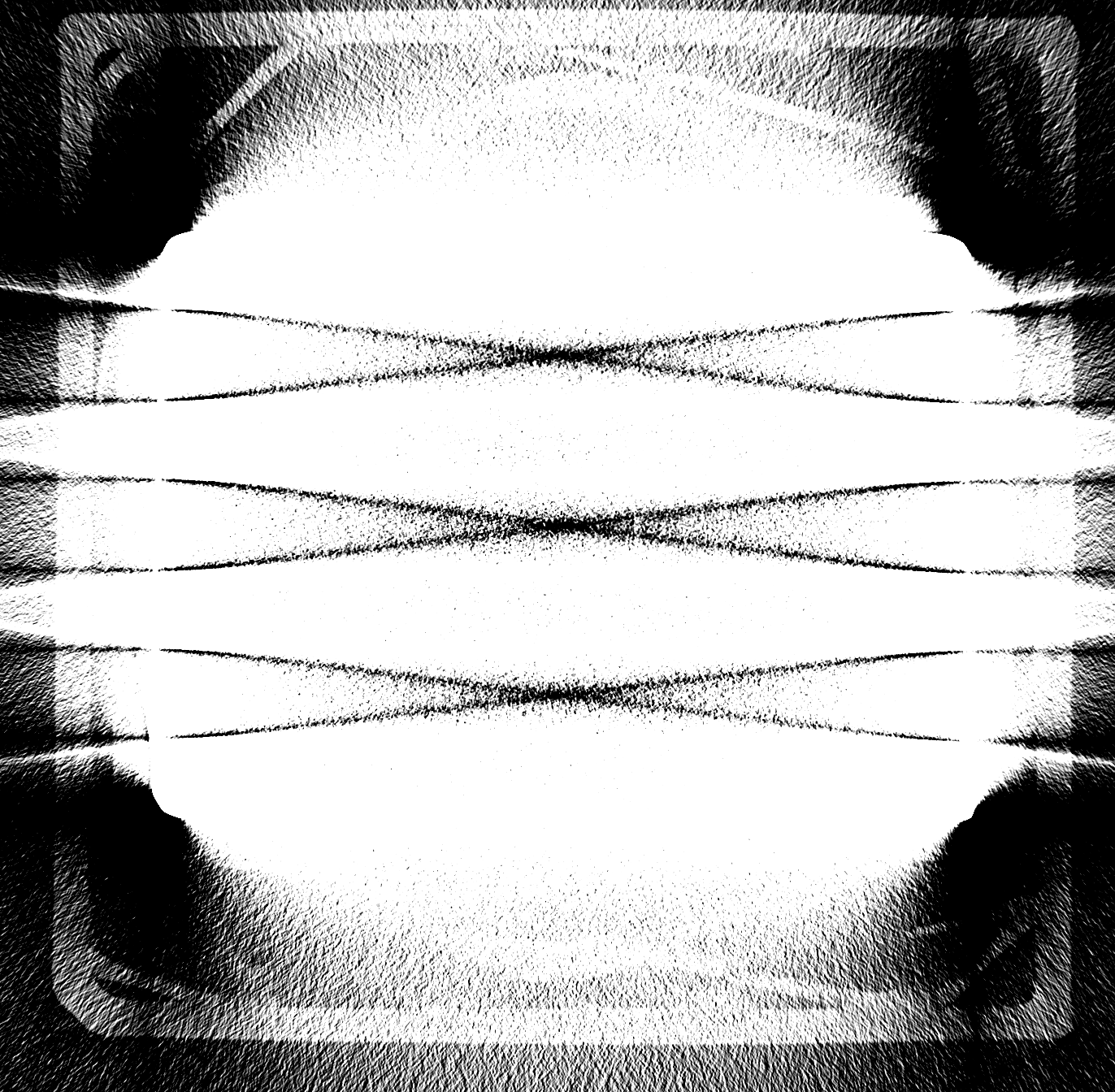}
\caption{Best reconstruction achieved with device's included
software}\label{fig:result-dev}
\end{figure}

\begin{pandoccrossrefsubfigures}

\subfloat[\ac{FBP}]{\includegraphics[width=6cm,height=\textheight]{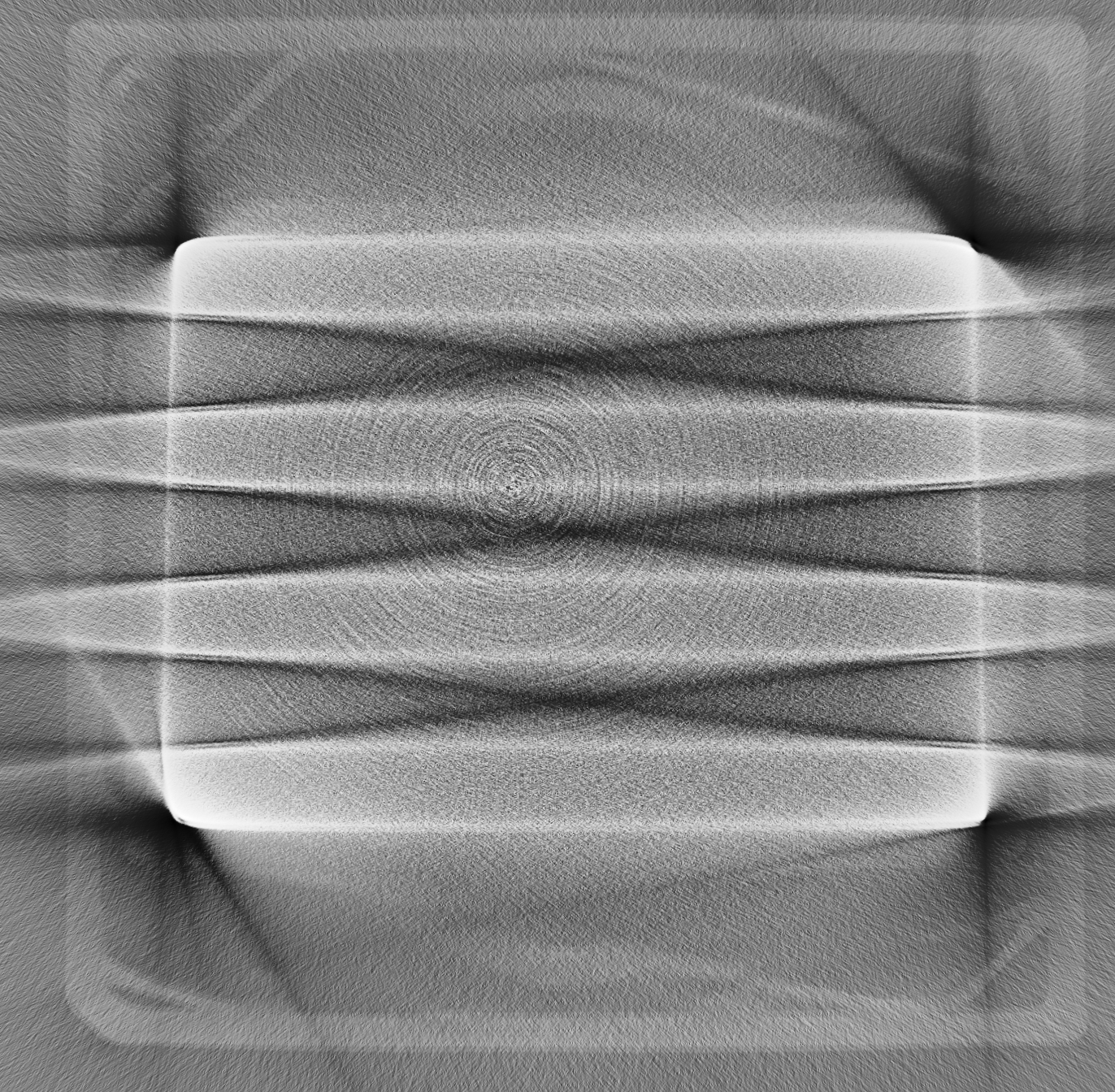}\label{fig:results1-fbp}}
\subfloat[\ac{CGLS}]{\includegraphics[width=6cm,height=\textheight]{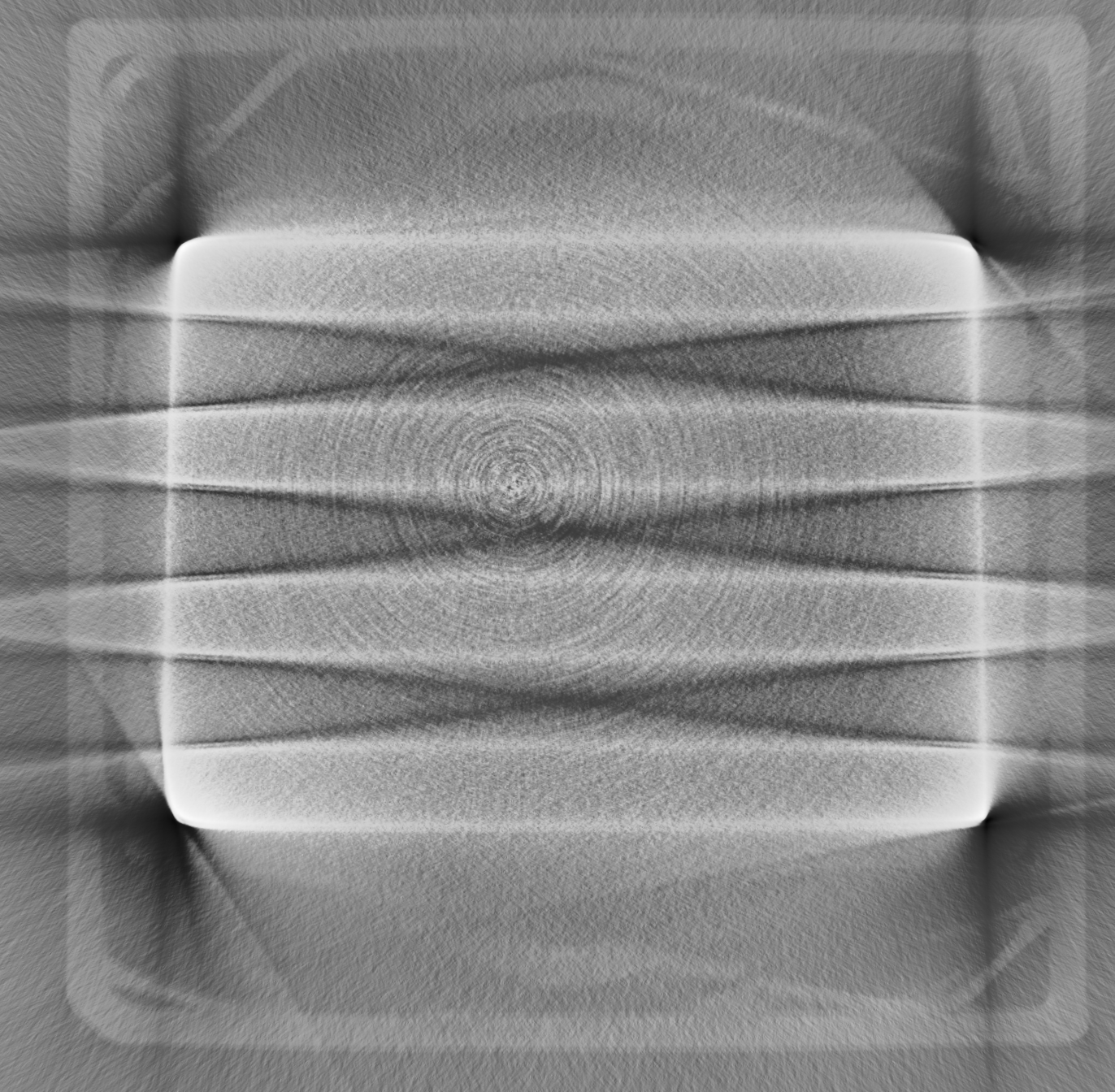}\label{fig:results1-cgls}}

\subfloat[\ac{FDK}
3-dimensional]{\includegraphics[width=6cm,height=\textheight]{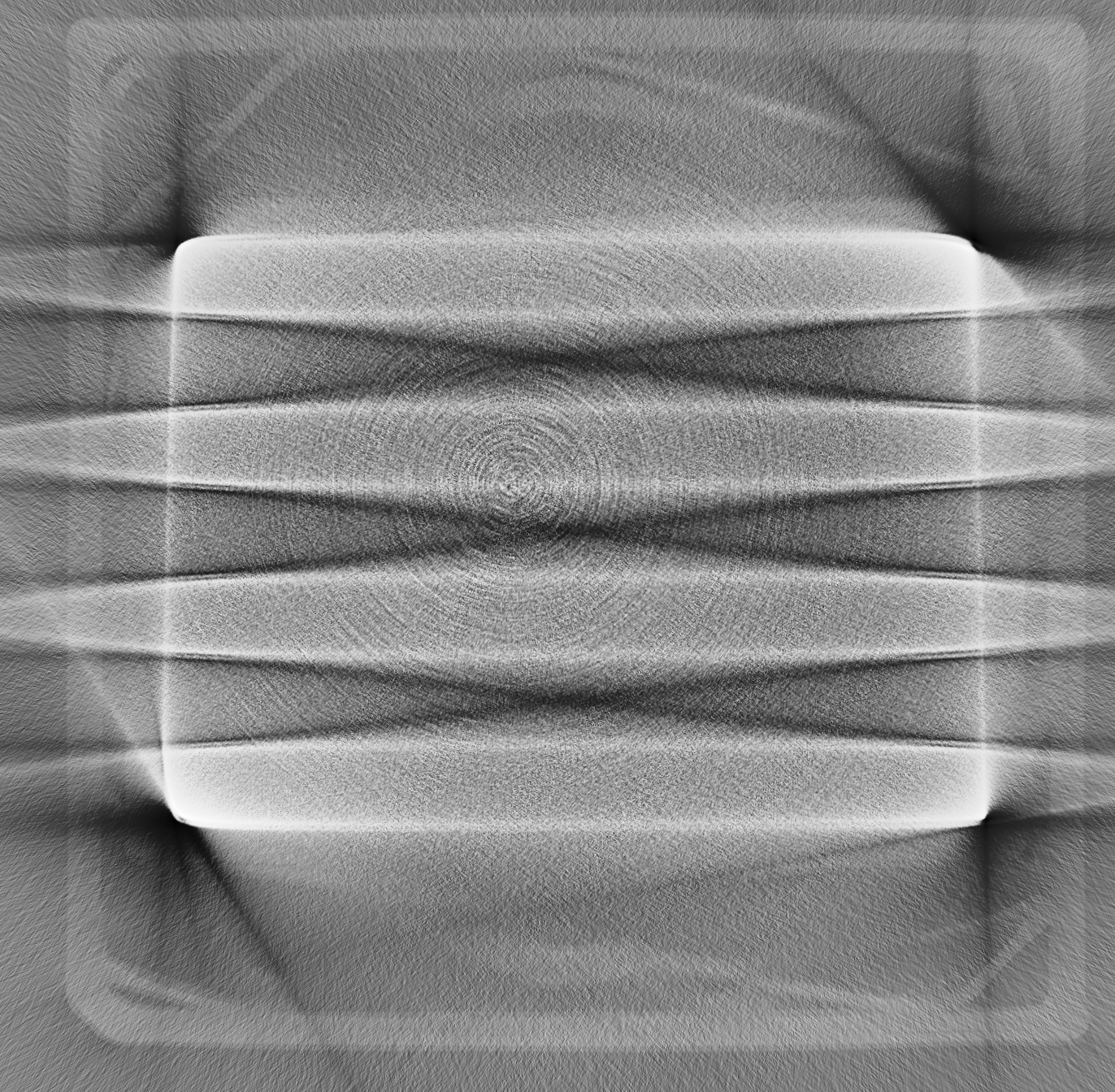}\label{fig:results1-fdk3d}}
\subfloat[\ac{SIRT}]{\includegraphics[width=6cm,height=\textheight]{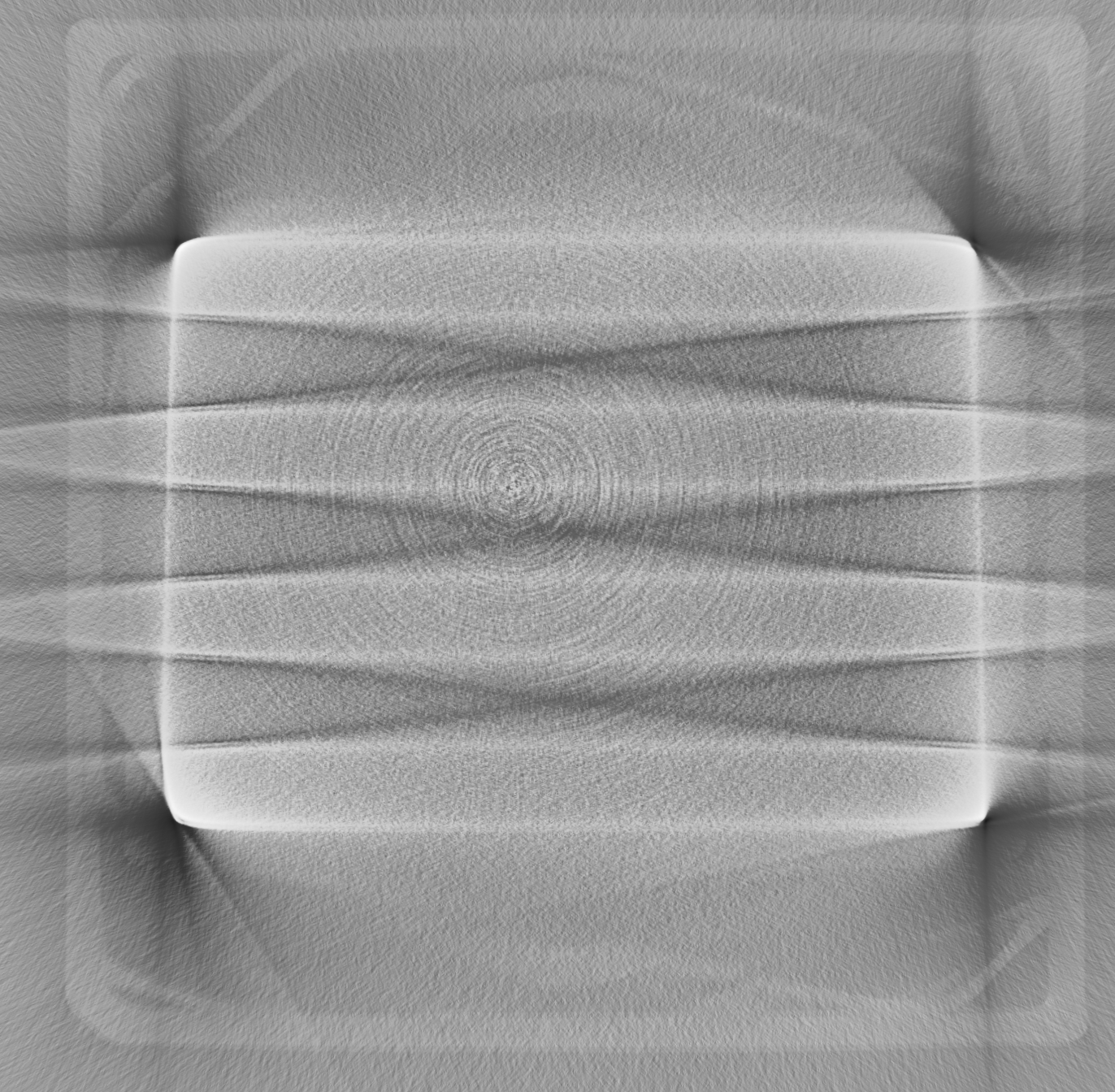}\label{fig:results1-sirt}}

\subfloat[\ac{SART}]{\includegraphics[width=6cm,height=\textheight]{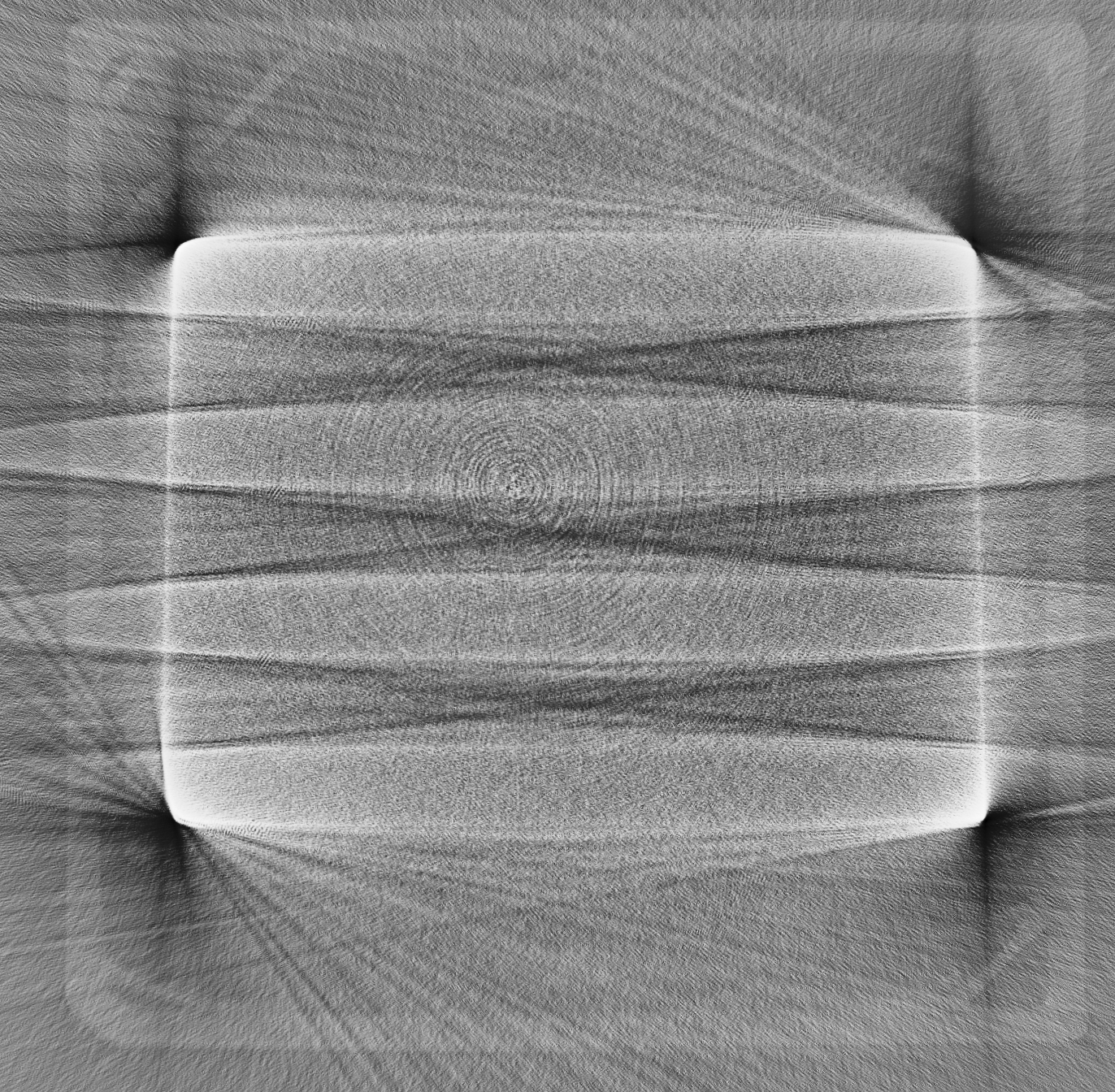}\label{fig:results1-sart}}
\subfloat[\ac{EM}]{\includegraphics[width=6cm,height=\textheight]{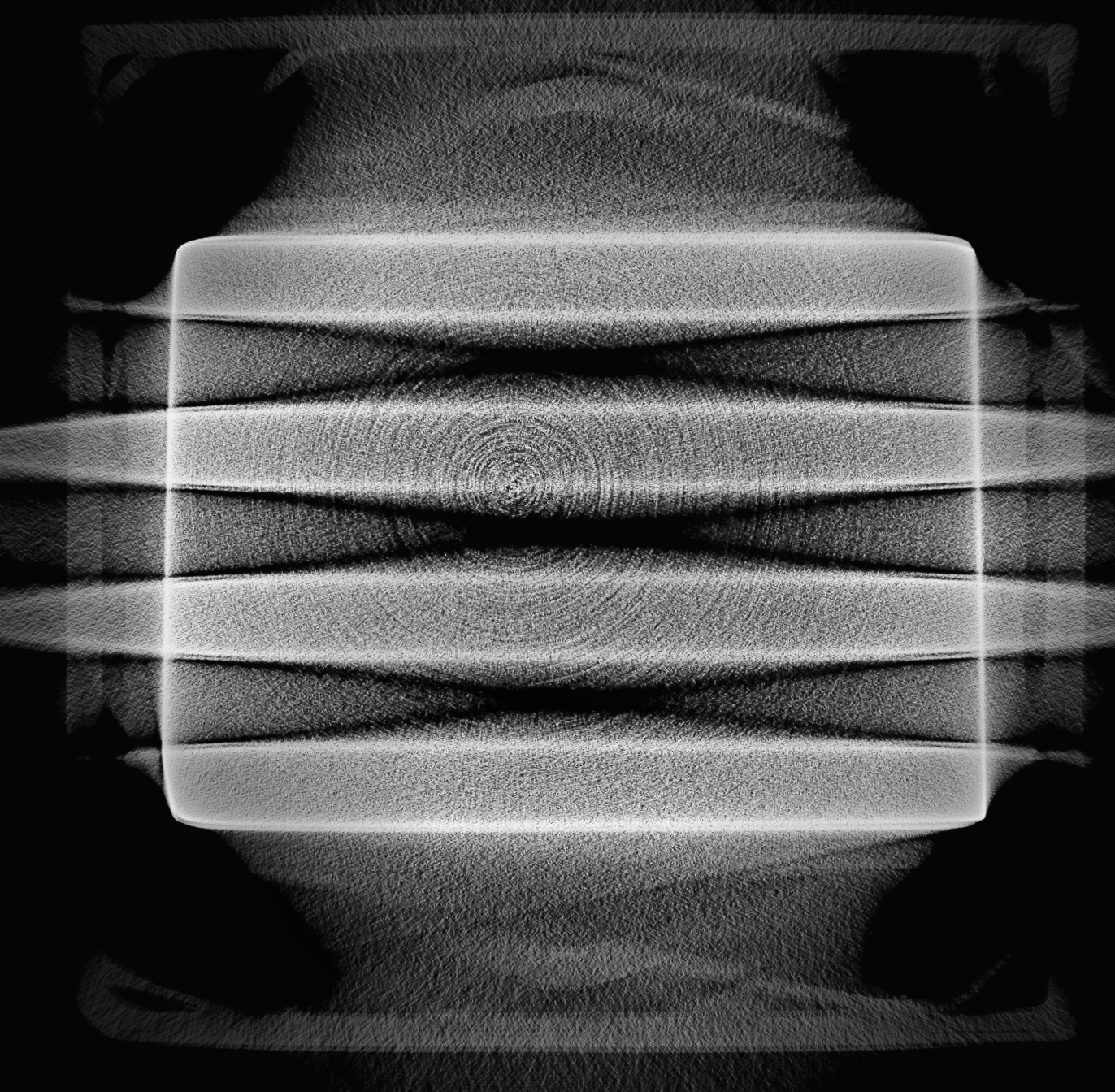}\label{fig:results1-em}}

\caption[{These are the post-processed results of the different testes
algorithms. Pictures have been adaptively normalized and range-reduced
to give a comparable viewing impression in this document.}]{These are
the post-processed results of the different testes algorithms. Pictures
have been adaptively normalized and range-reduced to give a comparable
viewing impression in this document.}

\label{fig:results1}

\end{pandoccrossrefsubfigures}

\subsection{Comparison}\label{comparison}

It is apparent from fig.~\ref{fig:results1}, that the undisclosed
algorithm (fig.~\ref{fig:result-dev}) in the manufacturer's software is,
in our case, performing worst even though it is tailored to the specific
device. Thus, it was decided to run also a \ac{FBP} reconstruction, as
this algorithm is very popular especially in more seasoned device
software and might be the base of the undisclosed algorithm. This way,
it was hoped, the comparison would be more objective and yield better
insight in the potential of the algorithms. The performance of \ac{FBP}
depends on the ability of the filter which is used and it turns out, the
default filter used by the ASTRA Toolbox is well suited to the tested
geometry.

It has to be noted, that almost all modern algorithms (means excluding
\ac{FBP}, \ac{FDK} and \ac{CGLS}) are able to receive a priori
information and/or weight maps which can greatly help to reduce artifact
generation and speed up reconstruction. There were no such information
provided in this test, as the establishment of a reliable weighting or
first-guess generator is not trivial and must be tailored to the
specific class of use-cases.

Further comparison yields, that \ac{SART} (fig.~\ref{fig:results1-sart})
generates noisy pictures and thus is deemed less favorable for the
typical geometry of electric propulsion ion optics. \ac{SIRT} generates
promising results quite fast even without a priori information, but
without this additional data it performs comparable to the classic
\ac{FBP} (given a good filter), \ac{FDK} and \ac{CGLS} algorithms.
\ac{EM} is very interesting as it produces high structural contrast, but
without a priori data or other methods to reduce the artifacts generate
by metal, the streaking severely impacts the usability of the algorithm.

\section{Conclusion}\label{conclusion}

The first conclusion of this endeavor is, that it can be very worthwhile
for experimenters to implement their own reconstruction solution instead
of using the manufacturer supplied software. The difference in quality,
depending on the implementation in the closed source software, can be
significant.

Secondly, while modern, especially iterative algorithms perform much
better in medical settings, the improvement is much smaller in the
setting of electric propulsion ion optics. The implementation of a
classic \ac{MAR} algorithms
{[}\citeproc{ref-changMetalArtifactReduction2012}{10}{]} was undertaken
but did not yield any usable results yet which could be presented here.
At the time of writing it is unclear, if this approach is feasible at
all with electric propulsion specimen. Further \ac{MAR} techniques have
yet to be investigated for applicability.

It can be assumed, that performance of reconstruction in modern
algorithms can be enhanced by introducing a priori data suited to this
use-case. Currently the \ac{SIRT} and \ac{EM} algorithms are deemed the
most favorable candidates for such an approach. Furthermore the
implementation of specific metal artifact reduction algorithms adapted
to electric propulsion diagnostics seems a viable endeavor.

\section*{Acknowledgments}\label{acknowledgments}
\addcontentsline{toc}{section}{Acknowledgments}

Part of this work was kindly sponsored by the \ac{GSTP} of the \ac{ESA}
{[}\citeproc{ref-esa400012491218NL2018}{11}{]}. Please refer to the
references section for details.

\section*{References}\label{references}
\addcontentsline{toc}{section}{References}

\phantomsection\label{refs}

\end{document}